\documentclass[final]{elsart5p}
\usepackage[dvips]{graphicx}
\usepackage{amsmath}
\usepackage{amssymb}
\usepackage{color}
\journal{JMMM}

\begin{document}

\begin{frontmatter}
\title{The dynamical response to the node defect in thermally activated remagnetization 
of magnetic dot array}
\author{P. Bal\'a\v{z}, D. Horv\'ath, M. Gmitra} 
\address{Department of Theoretical Physics and Astrophysics, Faculty of Science, P.J.\v{S}af\'arik University, Park Angelinum 9, 041 01 Ko\v sice, Slovak Republic}

  \begin{abstract}
    The influence of nonmagnetic central node defect on dynamical properties 
    of regular square-shaped $5 \times 5$ segment of magnetic dot array under 
    the thermal activation is investigated {\em via} computer simulations.
    Using stochastic Landau-Lifshitz-Gilbert equation we simulate hysteresis 
    and relaxation processes. The remarkable 
    quantitative and qualitative differences 
    between magnetic dot arrays with nonmagnetic 
    central node defect and magnetic 
    dot arrays without defects have been found. 
  \end{abstract}

  \begin{keyword}
    magnetic dot arrays \sep thermal activation \sep 
    stochastic Landau-Lifshitz-Gilbert equation \sep 
    magnetic hysteresis \sep magnetic relaxation
   \PACS 61.80.Az \sep 74.25.Fy \sep 74.72.Bk \sep 02.70.Bf 
  \end{keyword}
\end{frontmatter}


\section{Introduction}

  The term {\em Magnetic dot array} 
  (MDA)~\cite{Martin_2003,Srajer_2006,Ross_2002} 
  refers to the family of 
  nano-scaled monolayer structures consisting 
  of the identical magnetic nanoparticles, called dots, 
  which are periodically ordered on a non-magnetic substrate. 
  MDA concept is compelling, partly because of qualitatively 
  new properties, that essentially differ from those of the bulk materials. 
  Typical for MDA physics is an intricate collective behavior. 
  Its understanding may be valuable for condensed matter physics, 
  material science and nanoscience. The special properties of MDA systems 
  follow from the interplay between intra-dot and inter-dot interactions 
  as well as from the interplay of anomalously large surface 
  compared to the bulk magnetic 
  energy contributions~\cite{Horvath_2004a,Horvath_2004b}.
  The MDA properties are already 
  utilized in technological 
  applications concerning magnetic 
  field sensors~\cite{Duvail_1998,Thirion_2003,Hiebert_1997} 
  and reading heads of magnetic-disk data-storage devices \cite{Prinz_1995}.

  The technology of fabrication of MDAs~\cite{Chou_1997,McCleeland_2002,Ross_2001,Heyderman_2004} 
  has been perfected to an excellent degree in the past few years. 
  In practice, however, an occurrence of technological defects and 
  local irregularities has still a great influence on all of the 
  magnetic properties~\cite{Albrecht_2005}. 
  In this paper we study 
  how the imperfection in form of the single-dot vacancy affects 
  remagnetization of MDA under the assistance 
  of thermal activation.

  In the series of papers~\cite{Kayali_2003,Stamps_1998,Stamps_1999a,Stamps_1999b,Zhang_2005} 
  the authors focus on remagnetization of small segments 
  of arrays by 
  simulating elementary models where dots are treated as interacting 
  point dipoles. In Ref.\cite{Majchrak_2003} we draw the effect 
  of uniaxial anisotropy induced by the eccentrically placed node defect. 
  Later~\cite{Horvath_2006} we have investigated how the central defect 
  affects the quasi-static zero-temperature remagnetization of MDA. From 
  this study we know that: (i)~the square lattice seems 
  to be more sensitive to defect occurrence 
  than the triangular one; (ii)~zero-temperature differences of hysteresis 
  loops are not very pronounced despite of the remarkable changes 
  in the local arrangement of dots. In addition, the study has opened 
  question of defect influence on the relaxation modes. 
 
  In the present study the effect of nonmagnetic 
  central-node defect, in further referred 
  as {\it defect} only, is reconsidered.  
  However, the additional realistic factor 
  included here is the thermal activation. 
  In such case the statistical treatment of results 
  is necessary. Therefore, in order to discern defect 
  consequences, two distinct MDA 
  arrangements -{\em defect-free}~(DF) and 
  {\em defect-including}~(DI) are compared. 
 
 \section{Model}

  Because the statistical simulations need computational effort we preferred 
  use of elementary  model where each 
  dot is described by the point magnetic dipole.  
  This simplification is justified for monodomain 
  isotropic nearly spherical  
  ferromagnetic particles separated by a sufficient lattice spacing several times exceeding 
  a dot diameter. Additionally, we focus on the small MDA samples where dots 
  are placed on square $L \times L$ lattice. 
  The magnetic state of 
  $i$-th dot is described by the effective rescaled 
  3d magnetic moment 
  ${\bf m}_i$ normalized as $|{\bf m}_{i}|=1$.
  The inter-dot interactions 
  are assumed to be dipolar 
  and described by the effective field
  \begin{equation}
  {\bf h}_i^{\rm dip} = -  \sum_{j = 0, j \neq i}^{L \times L}
   \frac{{\bf m}_j r_{ij}^{2} -
   3 {\bf r}_{ij} ({\bf m}_{j} \cdot  {\bf r}_{ij})}
                     {r_{ij}^{5}}\,,
  \label{s_field_dip}
  \end{equation}
  where 
  ${\bf r}_{ij}$ is the distance between $i$-th and $j$-th 
  dot in lattice-spacing units $a$. 
  The field is measured 
  in the $H_{\rm 0} = V M_{\rm s} ( 4 \pi a^{3} )^{-1}$ 
  units including 
  the dot volume $V$ and saturated magnetization 
  $M_{\rm s}$.
  To study the DI arrangement we assume that defect 
  is represented by the zero magnetic moment. 
  The dynamics of magnetic moments is described by the 
  stochastic Landau-Lifshitz-Gilbert
  equation \cite{Scholz_2001}
  \begin{equation}
    \frac{{\rm d} {\bf m}_{i}}{{\rm d}\tau}
    = -
    {\bf m}_{i} \times {\bf h}_{i}^{\rm eff}
    -\alpha {\bf m}_{i}
    \times ({\bf m}_{i}\times {\bf h}_{i}^{\rm eff})\,,
  \label{sLLG}
  \end{equation}
  where $\alpha$ is the dimensionless damping parameter, 
  $\tau$ is the time in $t_{\rm 0}= 4 \pi a^{3} [ {\gamma ( 1+ \alpha^2 ) 
  V M_{\rm s}}]^{-1}$
  units and $\gamma$ is the gyromagnetic ratio. 
  The effective field
  ${\bf h}_{i}^{\rm eff}$ includes the dipolar field 
  ${\bf h}_i^{\rm dip}$, the external field ${\bf h}^{\rm ext}$, and
  the Langevin thermal field 
  ${\bf h}_{i}^{\rm th}$. Finally 
  ${\bf h}_{i}^{\rm eff} = {\bf h}_{i}^{\rm dip} +
  {\bf h}^{\rm ext} + {\bf h}_{i}^{\rm th}$.
  The random thermal field ${\bf h}_{i}^{\rm th}$ 
  is defined by averages  
 \cite{Garcia_1998}  
  \begin{subequations}
    \label{sprop}
    \begin{align}
      \langle h_{i, \xi}^{\rm th}(\tau) \rangle &= 0, \\
      \langle h_{i, \xi}^{\rm th}(\tau) 
     h_{j, \eta}^{\rm th}(\tau')\rangle &= 2 D \delta_{i j} \delta_{\xi \eta} \delta(\tau - \tau')\,,
    \end{align}
  \end{subequations}
  where $\xi, \eta \in \{x, y, z\}$  and $i$, $j$ are the site indexes;  
  $D$ is the noise amplitude.
  According to fluctuation-dissipation 
  relation~\cite{Scholz_2001,Garcia_1998}, 
  the factor $D$ is linked to the temperature $T$
  \begin{equation}
    D = \frac{\alpha}{1 + \alpha^{2}} \frac{T}{T_{0}}\,,
    \qquad
    \displaystyle
    T_{0} =   \frac{\mu_{0} V^2 M_{\rm s}^{2}}{  4 \pi k_{\rm B} a^{3}}\,,
  \label{disp}
  \end{equation}
  where $T_{0}$ is the characteristic temperature scale.

\section{Hysteresis}

  In all of our numerical experiments we set $L = 5$ for MDA. 
  The main empirical argument supporting 
  our choice $\alpha=0.1$ is that ferrite nanoparticles have 
  the mean value of $\alpha$ of such order~\cite{Fannin_2006}. 
  Let us first study the remagnetization in DF and DI systems in the time 
  varying external magnetic 
  field applied in parallel to one of 
  the main MDA axes 
  ${\bf h}^{\rm ext}(\tau) = 
  (h^{\rm ext}_{\rm x}(\tau), 0, 0)$. 
  For simulated geometry similarly as in the hypothetical experimental 
  setup the quantity 
  of interest is the magnetization projection 
  $\displaystyle M_{\rm x} = \frac{1}{L^{2}}\sum_{i = 1}^{L^{2}}
  {\bf m}_{i} \cdot {\bf e}_{\rm x}$. 
  The component $h^{\rm ext}_{\rm x}(\tau)$ 
  has been cycled within the bounds 
  $-h_{\rm max} < h^{\rm ext}_{\rm x}(\tau) < h_{\rm max}$.
  The simulation starts from the nearly saturated state with 
  $M_{\rm x} \simeq 1$ at $h_{\rm x}^{\rm ext} = h_{\rm max}$,  
  where $h_{\rm max}$ 
  is the bound chosen to keep 
  the system of moments saturated. 
  In the remagnetization 
  regime each time-integration step 
  $\Delta \tau= 10^{-2}$ is accompanied 
  by the unique change of the external field 
  $\Delta h_{\rm x}^{\rm ext}=\pm 10^{-6}$.  
  Both quantities define the sweeping rate 
  $v_{\rm h} =   (|\Delta h^{\rm ext}|/\Delta \tau) v_{0} 
  = 10^{-4}  v_{0}$,   where 
  $v_{0} = H_{0}/t_{0}$.  
  We observed that numerical results depend 
  on the sweeping rate. 
 
\begin{figure}[h]
    \centerline{\includegraphics[width=0.85\columnwidth, angle=0]{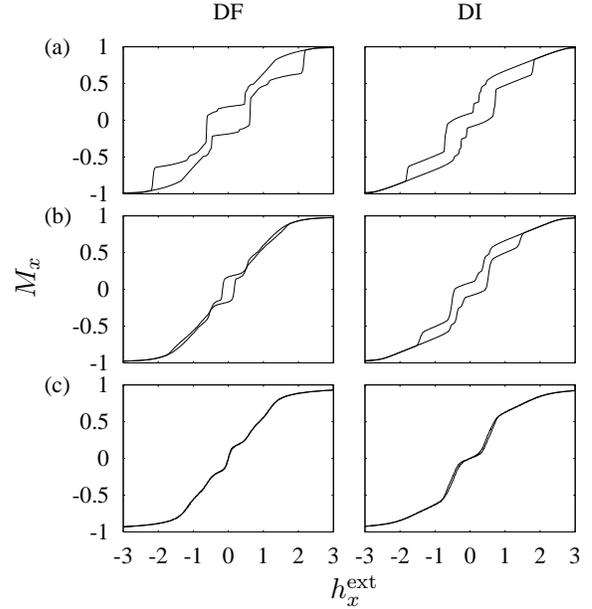}}
    \caption{The averaged hysteresis loops obtained for DF (left panel) 
             and DI (right panel) MDAs at temperatures 
             (a)~$T = 0.01 T_{0}$, 
             (b)~$T = 0.1 T_{0}$, 
             (c)~$T = 0.5 T_{0}$.}
    \label{mloops}
  \end{figure}

\begin{figure}[h]
   \centerline{\includegraphics[width=0.85\columnwidth, angle=0]{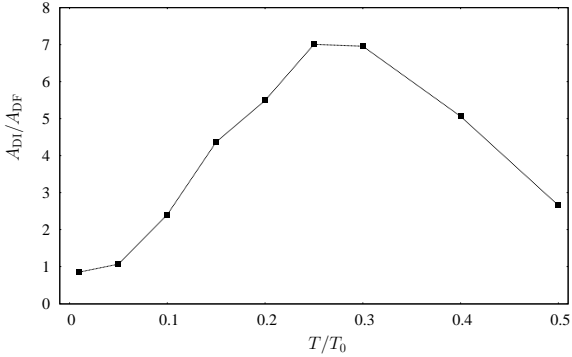}}
    \caption{The ratio of the areas of DI and DF hysteresis loops 
             as a function of the temperature.}
    \label{areas}
  \end{figure}

  Numerical integration of Eq. (\ref{sLLG})  has 
  been performed using stochastic 
  predictor-corrector Heun scheme \cite{Scholz_2001}. 
  This choice is justified by the fact that 
  in general, the statistical error 
  of the scheme can be made 
  arbitrarily small by averaging 
  over the number 
  of stochastic paths~\cite{Li_2003}. 
  We tested numerical stability 
  of scheme for two different integration steps. 
  In order to fix $v_{\rm h}$, 
  both $\Delta h_{\rm x}$ and 
  $\Delta \tau$ have been rescaled by 
  $1/2$. The test confirmed the invariance 
  of statistical results with respect to the rescaling. 
  Under the conditions 
  of thermal activation we recorded  
  and treated assembly 
  of $600$ independent loops. The treatment 
  assumes averaging of the magnetization 
  data conditioned by 
  $\Delta h_{\rm x}^{\rm ext}<0$ and 
  $\Delta h_{\rm x}^{\rm ext}>0$, respectively. 
  In Fig.\ref{mloops} we show averaged loops 
  constructed for $T=0.01 T_0$, $0.1 T_0$ and $0.5 T_0$. 
  We see that loops differ for DI MDA and DF MDA variants. 
  The difference confirms some anomalous impact 
  of the local assymetry of the couplings 
  broken by defect. Clearly, 
  by increasing the temperature the hysteresis 
  vanishes due to reduction of the impact of 
  irreversible processes due to configurations 
  separated by energy barriers.
  When assume standardly that the amount of free energy responsible 
  for the irreversible processes is proportional to the area $A$ 
  of hysteresis loop. 
  According Fig.\ref{areas} the thermally 
  induced hysteresis reveals maximum of the ratio 
  $A_{\rm DI}/A_{\rm DF}$ at $T=0.25 T_0$.  It is clear that 
  for sufficiently high temperatures the differences between 
  DI and DF systems vanish.
  Since the remagnetization process of MDAs is very inhomogeneous we analyzed the local mean magnetic hysteresis 
  for nearest neighborhood of central node, corners and middle nodes of MDA edges. The local loops differ 
  among each other. From their comparison it follows that defect affects not only hysteresis of its nearest 
  neighbor dots, but also hysteresis at corners and edges, that clearly follows from the long-range nature 
  of magnetostatic couplings. 

  One can expect that averaging eliminates some kind of the relevant defect-sensitive information.  
  Therefore, it seems valuable to pay attention to the fluctuations 
  around the averaged loops. To do this we analyzed the noise formed by returns 
  $\Delta M_{\rm x} =  M_{\rm x}(\tau +  100 \Delta \tau) - M_{\rm x}(\tau)$ 
  accumulated during reversals.  Also this data have been treated separately
  for $\Delta h_{\rm x}^{\rm ext} > 0$ and $\Delta h_{\rm x}^{\rm ext} < 0$ sweeps. 
  Evidently, the probability density functions of returns exhibit defect sensitivity. 
  They has been quantitatively characterized by the parameters of leptocurticity 
  $\langle  \Delta M_{\rm x}^{4} \rangle/3 \langle \Delta M_{\rm x}^{2} \rangle^{2}$  and 
  skewness $\langle |\Delta M_{\rm x}|^{3} \rangle/\langle |\Delta M_{\rm x}| \rangle^{3}$.
  Their numerical values are listed in Table.\ref{Tab:KS}.  
  \begin{table}[h]
  \begin{center}
    \begin{tabular}[h]{ccccc}
      \hline
      \multicolumn{1}{c}{ } &
      \multicolumn{2}{c}{leptocurticity} &
      \multicolumn{2}{c}{skewness} \\
       $T/T_{0}$ & DF & DI & DF & DI \\
      \hline
      \hline
      \hspace*{1em}0.01\hspace*{1em} & \hspace*{1em}66.52\hspace*{1em} & \hspace*{1em}41.31\hspace*{1em} &
             \hspace*{1em}2.28\hspace*{1em}  & \hspace*{1em}0.81\hspace*{1em} \\
      0.10 & ~4.41~ & ~6.91~ &
             ~0.05~ & ~0.06~ \\
      0.50 & ~1.93~ & ~1.93~ &
             ~0.04~ & ~0.03~ \\
      \hline
    \end{tabular}
    \caption{Statistical characteristics of the magnetization returns. The 
    leptocurticity and skewness obtained from fluctuations of hysteresis 
    at different temperatures.}
  \label{Tab:KS}
  \end{center}
  \end{table}
  The values clearly indicate that magnetization noise is strongly non-gaussian.  
  We see that temperature affects the characteristics in common manner, they decrease. 
  However, the remarkable differences between DF and DI systems exist again. 
 
\begin{figure}[h]
   \centerline{\includegraphics[width=0.85\columnwidth, angle=0]{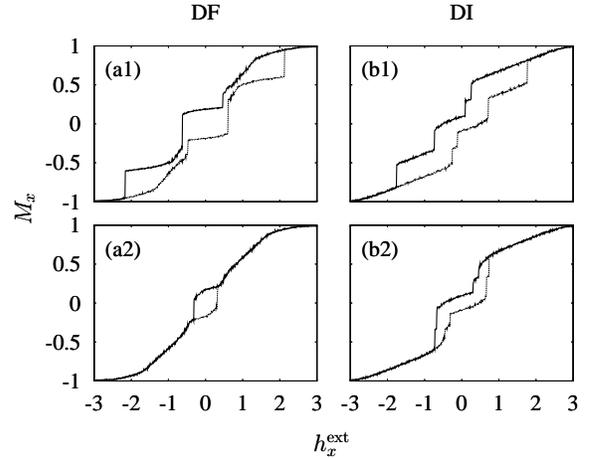}}
    \caption{The hysteresis loops for $T = 0.01 T_{0}$ revealed 
             by SOM network for DF and DI array. 
             The boldface branches belong to $\Delta h^{\rm ext} < 0$.}
    \label{koh}
  \end{figure}

  The reversal magnetization paths uncovered complex collective inner behavior of dot moments. 
  In addition the thermal fluctuations cause that single remagnetization events 
  qualitatively differ between each other. We observed that intervals with smoothly varying magnetization, 
  where the quasi-coherent rotation of moments prevails, are broken by the irreversible 
  jumps. To understand redundant but essential attributes of the reversal statistics 
  we exploited the abilities of artificial neural networks. The most efficient 
  for our purposes seems to be the usage of unsupervised self-organizing 
  maps (SOM)~\cite{Haykin1999, Reitzner_2004} that allow us to extract several 
  representative paths from the largely redundant assembly. In Fig.\ref{koh} we show 
  the representative paths ($T=0.01 T_0$) extracted by SOM. For certain inner parameters 
  of SOM the classification yields two representative groups of loops 
  (classification is done for DF and DI cases separately). 
  The quantitative output of this analysis is that the averaged loops from 
  Fig. \ref{mloops}(a) can be understood as a mixture of 89\% of (a1) constituent 
  with small addition of 11\% of (a2) in DF case, and 65\% of (b1) with 35\% contribution of (b2) 
  in DI case. This classification explains why the patterns (a1), (b1) are much more 
  similar to the averaged loops than less probable types (a2), (b2). 
  
\begin{figure}[h]
  \centerline{\includegraphics[width=0.95\columnwidth, angle=0]{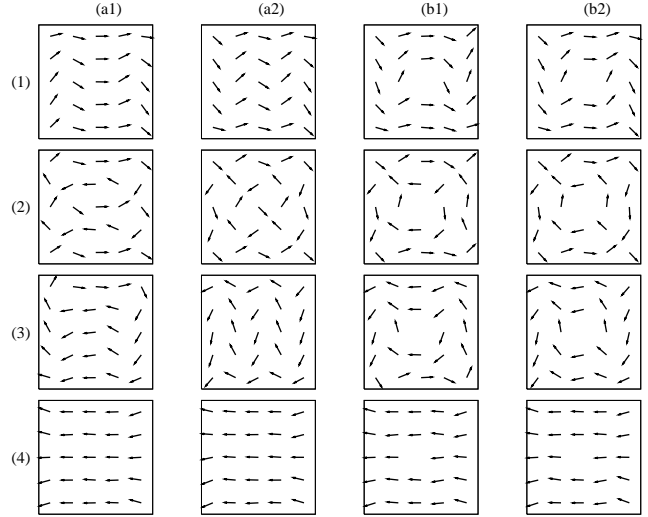}}
    \caption{The configuration of magnetic moments along the hysteresis path 
      that belongs to representative loop types obtained for decreasing 
     $h_{\rm x}^{\rm ext}$ equal to (1) 1.5, (2) 0.0, (3) -1.0, (4) -3.0.}
    \label{conf}
  \end{figure}

  By analyzing the revealed loops we conclude that two massive magnetization jumps contribute 
  to the reversal. Deeper insight to the mechanism of their nucleation is offered by the configuration 
  of moments. The sequence of snapshots that correspond to $\Delta h_{\rm x}^{\rm ext} < 0$
  regime is shown in Fig.\ref{conf}. We see that the defect evidently supports the nucleation 
  of the intermediate inter-dot {\em leaf}. Without defect the noncolinear antiferromagnetic 
  order is prefered. 

\section{Relaxation process}
  
\begin{figure}[h]
    \centerline{\includegraphics[width=0.85\columnwidth, angle=0]{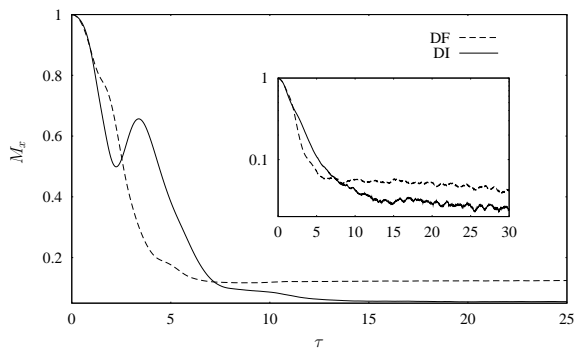}}
    \caption{The averaged magnetization relaxation obtained for DF and DI systems 
      at temperatures $T = 0.01 T_{0}$. Inset corresponds to $T = 0.5 T_{0}$. 
      The final curves are averages obtained from 1000 
      stochastic relaxation events.}
    \label{rel}
  \end{figure}

  The additional principle of defect detection is provided by relaxation process 
  in zero external field starting from $M_{\rm x} = 1$ state. 
  The relaxation curves for DF and DI systems are plotted in Fig.\ref{rel}. 
  For $T = 0.01 T_{0}$  the qualitative differences in the relaxation 
  of magnetization become clear. The most anomalous aspect of relaxation 
  of DI is the pronounced peak. However, DF MDA does not exhibit this feature. 
  The configurations show 
  that in DF case the pair of antiparallel vortices is formed in contrast to inter-dot 
  flower observed in DI case. As for the hysteresis, when the energy barriers 
  are over bridged by thermal fluctuations, 
  the transition to relaxation mode without peak is expected. 
  This scenario is confirmed 
  by the inset of Fig. \ref{rel}, 
  where the nearly exponential relaxation occurs without 
  substantial marks of defect sensitivity. 
 

\section{Conclusion and discussion}

  The simulation statistical study discusses the theoretical 
  grounds for defect detection in MDA. 
  It has been shown, that that the presence of 
  non-magnetic central node defect in a a small segment of MDA has non-negligible influence 
  on dynamical properties of MDA, concretely magnetic hysteresis and relaxation.
  Further, it shows that the including the thermal activation is not only the additional factor 
  of realistic simulation, but also the factor that can substantially enhance the chance to identify defect. 
  
  Finally, we are concluding with the believe that our present results will affect 
  the experimental claims in this direction.
  
  \vspace{4mm}
  
  The authors would like to thank for financial support through grants 
  VEGA 1/2009/05, APVT-51-052702, APVV-LPP-0030-06.

\end{document}